\documentclass[12pt]{iopart}
\usepackage[dvips]{graphicx}
\usepackage{epsfig}
\newcommand\beq{\begin{equation}}
\newcommand\eeq{\end{equation}}
\newcommand\bea{\begin{eqnarray}}
\newcommand\eea{\end{eqnarray}}

\begin{document}

\title[Double ionization in strong fields]%
{Classical threshold behaviour in a 1+1-dimensional model 
for double ionization in
strong fields}

\author{Bruno Eckhardt\dag and Krzysztof Sacha\ddag\ 
}

\address{\dag\ Fachbereich Physik, Philipps-Universit\"at Marburg,
D-35032 Marburg, Germany}

\address{\ddag\ Instytut Fizyki im. Mariana Smoluchowskiego,
Uniwersytet Jagiello\'nski, ul. Reymonta 4,
PL-30-059 Krak\'ow, Poland}

\begin{abstract}
Building on insights about the classical pathways to double
ionization in strong laser fields we here propose a 1+1-dimensional
model that captures essentials of the 3-d potential landscape and 
allows efficient studies of the process in reduced dimensions.
The reduction to one degree of freedom for each electron is
obtained by confining their motion to lines at an angle
of $\pi/6$ with respect to the field axis; the justification for
this choice is that upon variation of the electric field the
Stark saddles move along the same lines. In this way we obtain a
low-dimensional representation in which symmetric electron escape
is possible. As a demonstration of the advantages of this model
we confirm numerically the equivalent of the Wannier threshold behaviour
for constant electric field and identify several classes of 
trajectories that can contribute to the ionization process.
\end{abstract}


{\bf Introduction.}
Analyzing the dynamics of two-electron systems remains a challenge 
because of the high dimensionality of the configuration space, especially
when highly excited states are involved, as in the case of strong field
double ionization. Taylor group managed to obtain results for a full 
3+3-dimensional representation, albeit mostly at short wave length
\cite{taylor1,taylor2}. 
1+1-dimensional models with both electrons aligned along one axis 
\cite{align1,align2,align3,align4} are attractive (as reflected in a 
large number of publications based on this model), 
but physically flawed since the electron repulsion prevents an escape 
in the experimentally observed subspace of equal momenta
\cite{weber00n,wechenbrock01,feuerstein01,moshammer02,moshammer03,dorner,moshammer5}. 
Among the
alternatives that have been considered are models where the center of mass
of the electrons is confined to move along the axis \cite{becker6}: they might be classified
as 1.5+1.5 dimensional, since the position of an electron along the field axis 
qualifies as a full degree of freedom and the perpendicular one,
being shared with the other electron, as half a degree of freedom. 
It is our aim here to present and analyze a model with further
constraints on the dynamics; the model nevertheless captures several
essential features of the full dynamics and allows efficient
simulations.

The model is suggested by our previous analysis of the classical pathways
to non-sequential double ionization in strong laser fields
\cite{eckhardt01pra1,eckhardt01,eckhardt01epl}. 
We found that
the observed symmetric escape can be explained by the escape over a
symmetrically arranged saddle. As the field changes, the
saddle moves along lines that keep a constant
angle with respect to the polarization axis.
Confining the electrons to run along 1-d tracks that 
pass through the saddles gives a model for 
1+1 dimensional electron dynamics that
has a potential landscape topology very similar
to that for electrons in the full 3-d case.
While the potential turns out to be very similar
to the so called aligned-electron model \cite{align1,align2,align3,align4}, 
it has one significant difference: since the
electron tracks separate as they move away from the
nucleus electron repulsion diminishes as the electrons
go out. Thus, the diagonal where the two electron
coordinates are equal, is accessible and not, 
as in the aligned model, suppressed by
electron repulsion. This allows to mimic a key
feature of the observed double ionization dynamics,
namely the preference for symmetric electron escape
\cite{weber00n,wechenbrock01,feuerstein01,moshammer02,moshammer03,dorner,moshammer5}.

In the present publication we analyze the model classically for 
the case of a static electric field. The static field forms 
a local maximum in the potential energy (that corresponds to the saddle 
in the 3-d case \cite{eckhardt01pra1}) close to which simultaneous
escape of electrons takes place. If the energy of the system 
equals the energy of the stationary point only purely symmetric 
electron motion leads to the simultaneous escape. For higher energy
some deviations from the symmetric motion are allowed. The stability 
analysis of the saddle point in the full 3-d case allowed us to 
predict the dependence of the classical cross section for the process 
on energy close to the threshold \cite{eckhardt01epl}, i.e. 
to obtain the Wannier threshold 
law  \cite{wannier53,rau84,rost98}
in the presence of the external field. However, the results of the 
local analysis were difficult to verify in 3-d numerical simulations. 
The key features of the 3-d potential are present in our 1+1 
dimensional model that allows us to obtain also the threshold law and 
moreover verify it in numerical simulations.

{\bf The model.}
We begin with a reminder of the saddle configurations in the symmetric 
subspace in the 3-d case \cite{eckhardt01pra1,eckhardt01,eckhardt01epl}.  
Assume the external electric field points in the $z$-direction and 
the electrons are labeled $i=1,2$ with positions 
${\bf r}_i=(x_i,y_i,z_i)$. Then the Hamiltonian, in atomic units, is
\beq
H = \sum_{i=1}^2 \left\{ \frac{{\bf p}_i^2}{2}- \frac{2}{|{\bf r}_i|}
- F z_i\right\}
+ \frac{1}{|{\bf r}_1-{\bf r}_2|}\,.
\label{ham}
\eeq
Let the electrons be moving in the $x$-$z$-plane,  
symmetric with respect to the field axis. Then their positions
are $(x,0,z)$ and $(-x,0,z)$ and the potential energy becomes
\beq
V=-\frac{4}{\sqrt{x^2+z^2}}
+\frac{1}{2|x|} - 2 F z.
\label{hcNv}
\eeq
This potential energy has a saddle at
$x_s=3^{1/4}/(2\sqrt{F})$, $z_s=3^{3/4}/(2\sqrt{F})$, 
with energy $-3^{3/4} 2 \sqrt{F}$.
Note that if we allow for variation of the field strength 
(like, e.g., in a laser field) the saddle moves along lines 
with $z_s/x_s=\sqrt{3}=$const. Because simultaneous electron
escape takes place in the vicinity of the saddle, the idea is 
then to assume that 
the electrons move in a plane, $y_i=0$, with coordinates 
constrained exactly by this relation, $z_i^2=3 x_i^2$. 
This leaves only one degree of freedom for each electron,
called $r_1$ and $r_2$:
\beq
x_1=\frac{1}{2} r_1\quad
z_1=\frac{\sqrt{3}}{2} r_1
\qquad
\mbox{and}
\qquad
x_2=-\frac{1}{2} r_2\quad
z_2=\frac{\sqrt{3}}{2} r_2\,.
\eeq
The Hamiltonian becomes
\beq
H = \frac{p_1^2+p_2^2}{2}-\frac{2}{|r_1|}-\frac{2}{|r_2|}+
\frac{1}{\sqrt{(r_1-r_2)^2+r_1r_2}}-
\frac{F\sqrt{3}}{2}(r_1+r_2).\;
\label{fin}
\eeq
The Hamiltonian (\ref{fin}) defines our 1+1 dimensional model and in following 
we restrict ourselves to description of the two electron system within 
this model.

\begin{figure}
\centering\epsfig{file=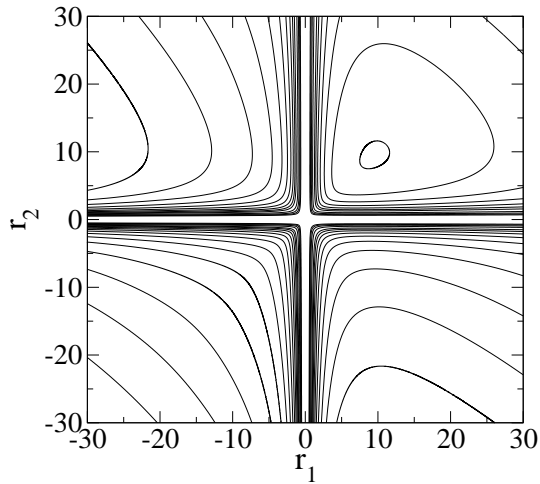,width=10.cm}
\caption[]{\label{p0.02}
Equipotential contours of the potential energy in (\ref{fin}) 
for electric field amplitude $F=0.02$.
}
\end{figure}

When the external field is present the potential energy in (\ref{fin}) 
has a saddle located at 
\beq
r_1=r_2=r_s=\frac{3^{1/4}}{\sqrt{F}}\,,
\label{rs}
\eeq
with energy
\beq
V_s=-3^{3/4} 2 \sqrt{F}.
\label{vs}
\eeq
The equipotential contours for $F=0.02$ are plotted in Fig.~\ref{p0.02};
the stationary point near $r_1=r_2\approx 9.3$ is clearly visible.  
There is a symmetric subspace in the full phase space of the 1+1 dimensional 
system and the stationary point lives in this subspace. That is, if initial 
conditions 
are chosen symmetrically, i.e. $r_1=r_2$ and $p_1=p_2$, the electron motion 
remains symmetric in time evolution. Trajectories that pass close to 
the stationary point and sufficiently close to the symmetric subspace 
lead to simultaneous electron escapes.

{\bf Wannier threshold law.}
The potential energy in (\ref{fin}) possesses a stationary point 
located at (\ref{rs}), i.e. in the symmetric subspace of the system.
The stability analysis of the stationary point in the full phase space 
allows us to determine the behaviour of the cross-section for 
simultaneous electron escape close to the threshold \cite{eckhardt01epl}. 
There are two Lyapunov exponents of the point. The first, 
\beq
\lambda_r=3^{1/8} F^{3/4} \,, 
\eeq
with the eigenvector in the symmetric subspace, 
corresponds to a simultaneous motion of the electrons 
in the same direction away from the
nucleus. Borrowing terminology from chemical reactions, 
we call this subspace the reaction coordinate. Because
of the repulsion between electrons, there is an additional unstable 
direction, with Lyapunov exponent
\beq
\lambda_\perp= \sqrt{\frac{11}{6}} 3^{1/8} F^{3/4} \,,
\eeq
which enters in the threshold law. 
If the initial energy of the system equals the stationary point energy
only a trajectory living in the symmetric subspace can lead to
a simultaneous escape of both electrons. This reduces the dimensionality of the
problem and the cross-section vanishes. For higher energy some deviations from
the symmetric motion are possible, giving a finite volume of 
initial conditions and a non-vanishing cross-section.
The dependence of the cross-section on energy $\sigma(E)$ 
close to the stationary point 
energy $V_s$ can be obtained in the spirit of the Wannier analysis
\cite{eckhardt01epl,wannier53,rau84,rost98,rost01phe}, resulting in
\beq
\sigma(E) \sim (E-V_s)^\mu,
\label{thresh}
\eeq
with an exponent
\beq
\mu=\frac{\lambda_\perp}{\lambda_r}=\sqrt{\frac{11}{6}}\approx 1.354 \,.
\label{cse}
\eeq
This is larger than the corresponding exponent
in the full 3-d case where, for a doubly charged remaining ion, 
the exponent is $1.292$ \cite{eckhardt01epl}.
The cross section is larger if the exponent is smaller, i.e. if 
the saddle is crossed more quickly (larger $\lambda_r$)
or if the differences from the symmetric motion grow more slowly
(smaller $\lambda_\perp$).
This cross section exponent is an additional characteristic
of the double ionization process.

{\bf Numerical results.}
The derivation of the cross section exponent (\ref{cse}) for the simultaneous
electron escape is based on the local analysis of the stationary point of the 
system. One may wonder if the higher order terms can modify the behaviour 
of the cross section. To test the results of the local analysis we can perform
numerical simulations of the process. In the 3-d case considered in 
Ref.~\cite{eckhardt01epl} that was quite difficult due to high dimensionality 
of the problem. Here, we deal effectively with the three dimensional phase 
space and the numerical simulations become feasible. 

\begin{figure}
\centering\epsfig{file=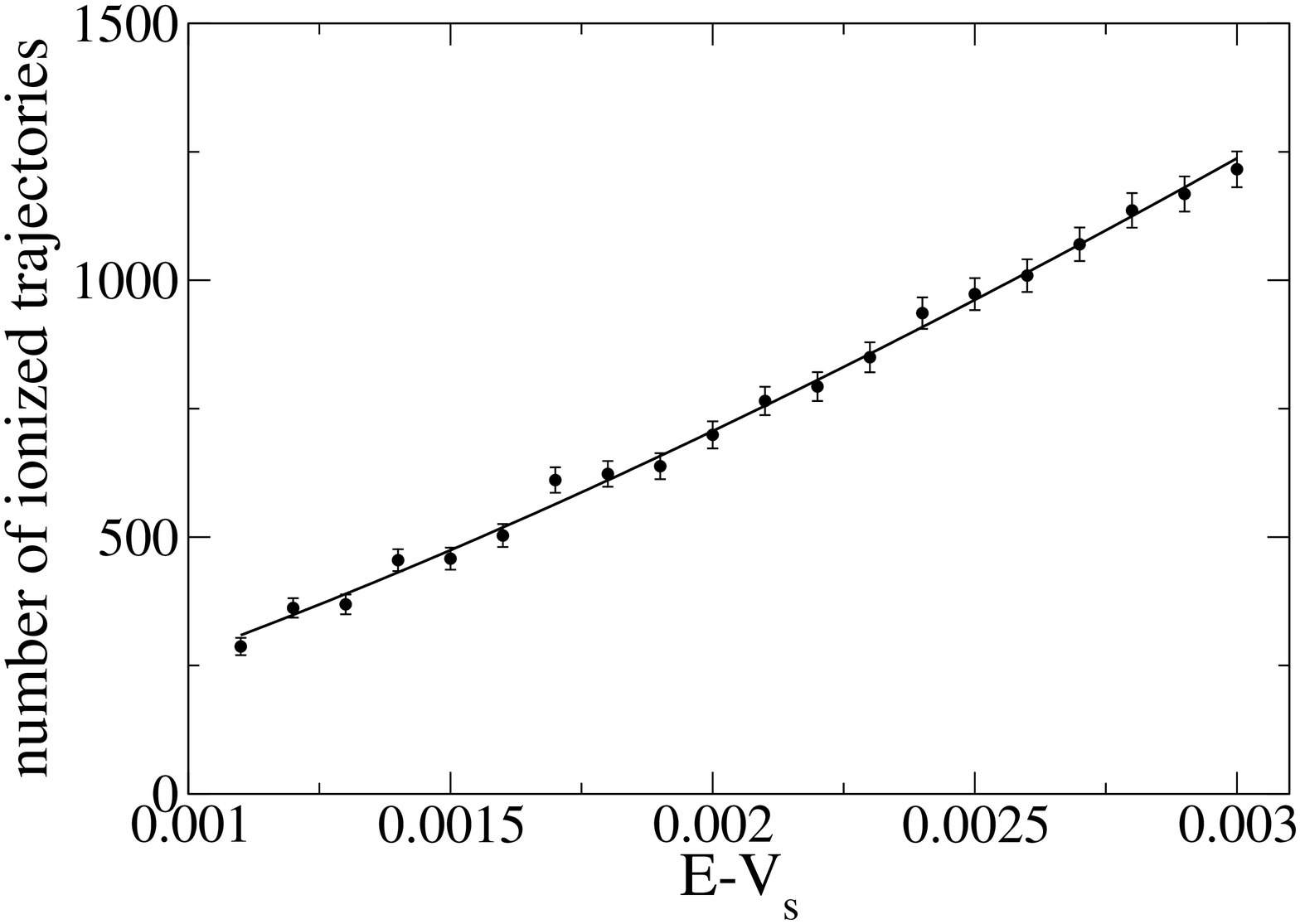,width=10.cm}
\caption[]{\label{numerics}
Number of ionized trajectories (circles with statistical error bars)
versus energy above the threshold energy, i.e. the energy of the maximum
of the potential (\ref{vc}), corresponding to $F=0.02$ ($1.5\times 10^6$
trajectories have been integrated to obtain each point).
Solid line is the fit of (\ref{thresh}) to 
the numerical data that results in $\mu=1.383\pm0.031$.
}
\end{figure}

In order to avoid problems with the Coulomb singularities,
we can add cut-offs in the denominator, so that 
the potential becomes
\beq
V = -\frac{2}{\sqrt{r_1^2+1}}-\frac{2}{\sqrt{r_2^2+1}}+
\frac{1}{\sqrt{(r_1-r_2)^2+r_1r_2+1}}\,-\frac{F\sqrt{3}}{2}(r_1+r_2).
\label{vc}
\eeq
The cut-offs change slightly the prediction for the threshold exponent
which now equals $\mu= 1.357$ (for $F=0.02$). We have run trajectories with 
initial conditions chosen microcanonically for different energies $E$ above 
the threshold energy $V_s$ but with an additional requirement that they have 
to lie on the surface $r_1+r_2=0$. We deal with an open system and the 
condition $r_1+r_2=0$ ensures that the trajectories start in the vicinity of 
the nucleus. In Fig.~\ref{numerics} number of
trajectories leading to simultaneous electron escape together with
the numerical fit of the function (\ref{thresh}) versus energy are shown. 
The fitted value of the cross section exponent is $1.383\pm0.031$ what 
agrees with the theoretical prediction. 

\begin{figure}
\centering\epsfig{file=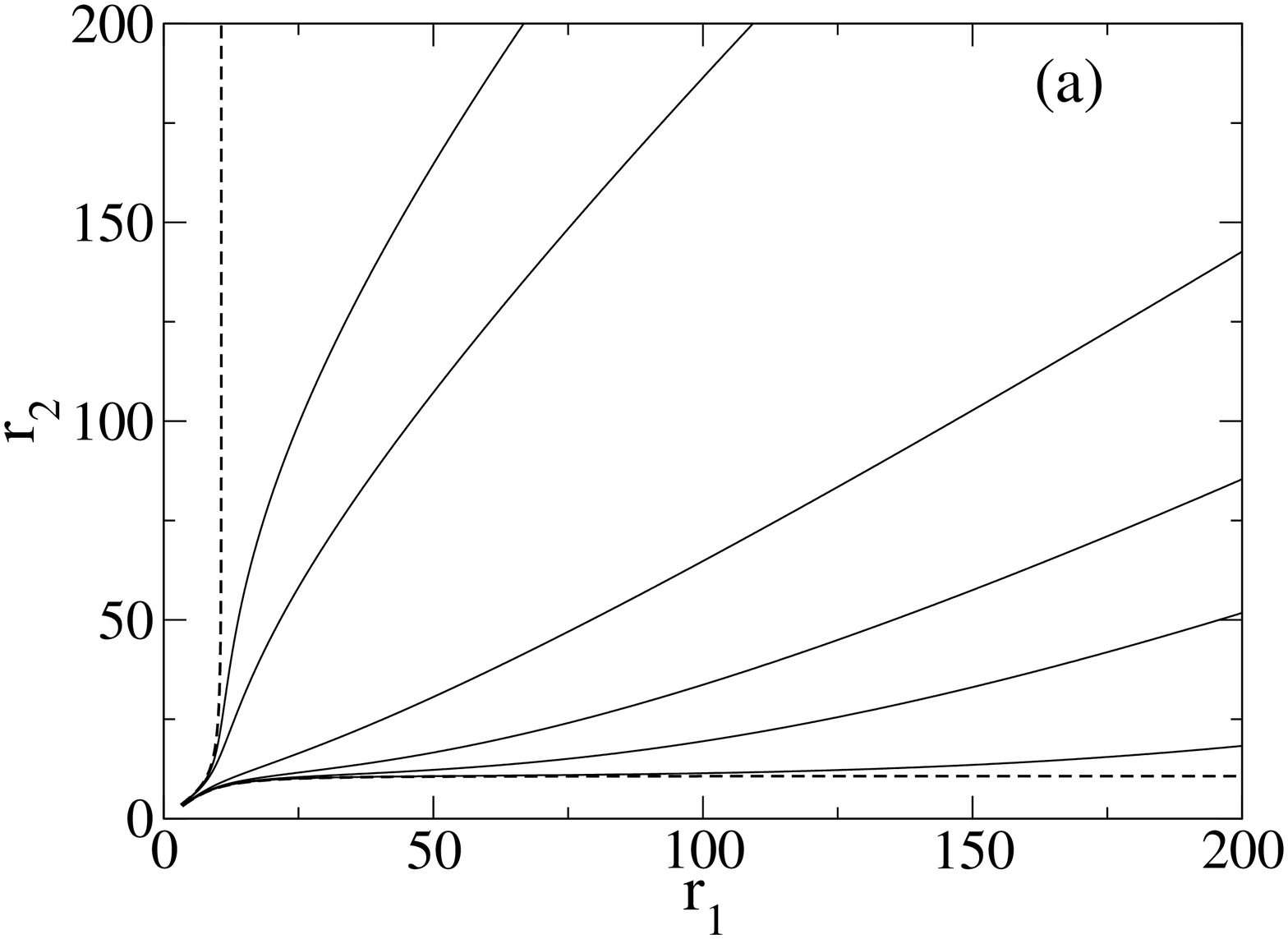,width=8cm}\epsfig{file=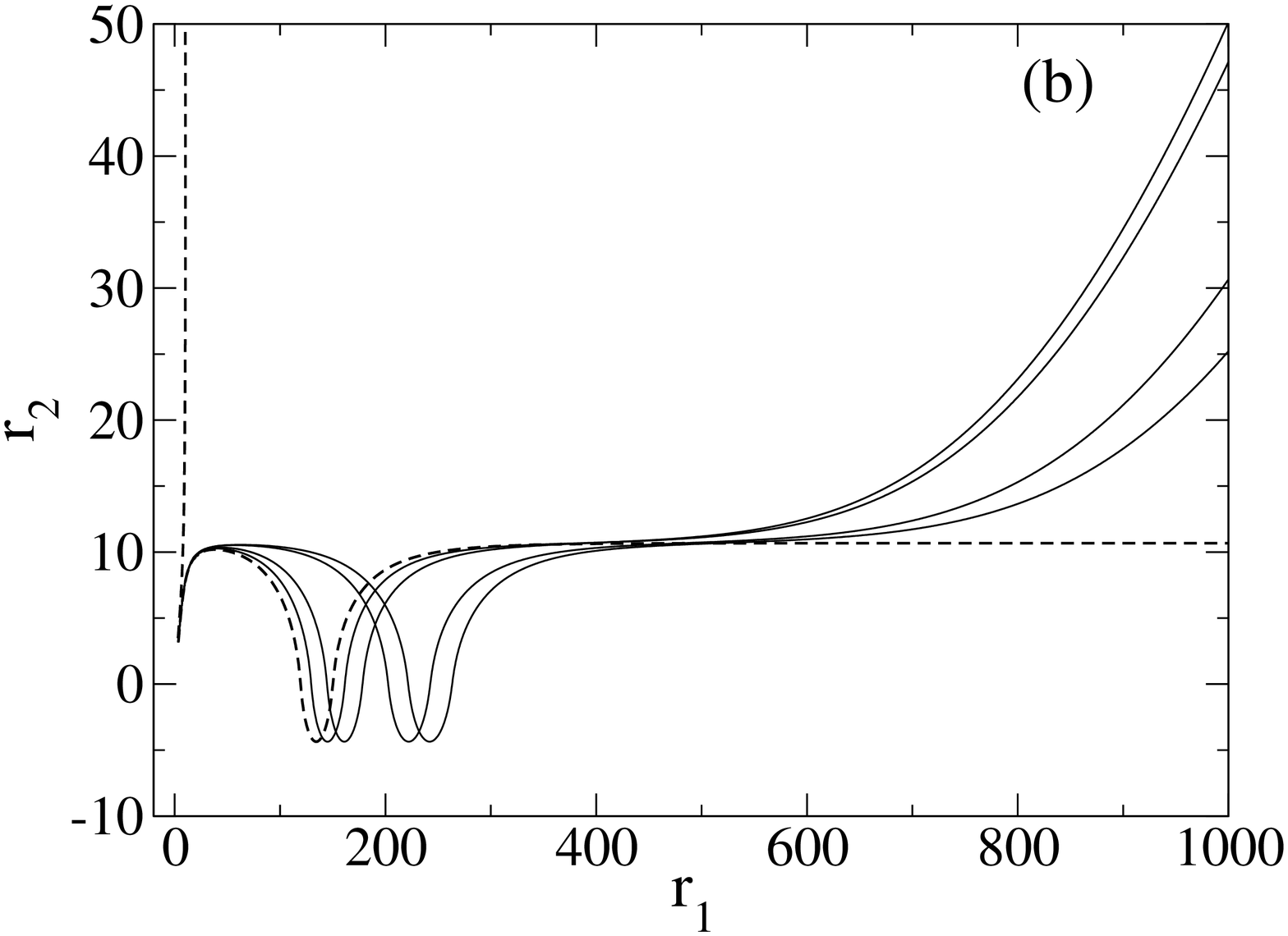,width=8cm}
\caption[]{\label{traj}
Panel (a): trajectories (dashed lines) belonging to the separatrix 
that divides the phase space into areas of simultaneous electron escape 
and double escape with a single revisit of the nucleus by one of the electrons;
solid lines are examples of trajectories leading to simultaneous electron
escape. 
Panel (b): trajectories (dashed lines) belonging to the separatrix 
that divides double ionization trajectories with single 
and double revisit of the nucleus;
solid lines are examples of trajectories leading to double ionization
with a single revisit of the nucleus by one of the electrons.  
All trajectories correspond to: $F=0.02$, energy $E=V_s+0.1$, 
symmetric initial momenta $p_1=p_2$ and positions lying on a circle, i.e. 
$r_1=(r_s/2)\cos\alpha$ and $r_1=(r_s/2)\sin\alpha$.
}
\end{figure}

In Fig.~\ref{numerics} we have included only trajectories leading to 
simultaneous escape of both electron. However, it is possible that the first
electron ionizes by passing close to the stationary point while the other 
one is returned and after revisiting the nucleus ionizes too. Actually 
there is a separatrix that divides the phase space into areas corresponding 
to simultaneous escapes and escapes with a single revisit of the nucleus. 
The separatrix consists of trajectories where the first electron escapes 
while the other one approaches a single electron Stark saddle and looses its
kinetic energy. 
Finally, when the first electron is gone, the other electron remains at 
the single electron saddle with no kinetic energy. In Fig.~\ref{traj} there are
examples of ionized trajectories and trajectories belonging to the separatrix. 
All trajectories in Fig.~\ref{traj} correspond to: $F=0.02$, energy $E=V_s+0.1$, 
symmetric initial momenta $p_1=p_2$ and positions lying on a circle, i.e. 
$r_1=(r_s/2)\cos\alpha$ and $r_1=(r_s/2)\sin\alpha$. With these conditions 
if $|\pi/4-\alpha|$ is smaller than about $0.0154\pi$ simultaneous 
escape takes place. 
For greater $|\pi/4-\alpha|$ but smaller than about $0.0155\pi$ the second 
electron revisits the
nucleus and then escapes. At $|\pi/4-\alpha|\approx0.0155\pi$ another
separatrix appears that separates double ionization trajectories with single 
and double revisit of the nucleus. This separatrix (see Fig.~\ref{traj}) 
consists of trajectories where the second electron after revisiting the 
nucleus approaches the single electron Stark saddle loosing its kinetic 
energy so that finally it stands on the saddle. With further increase 
of $|\pi/4-\alpha|$, number of 
revisits increases quickly until the returning electron has too small energy 
to cross the single electron saddle even if the first electron is gone.

{\bf Conclusions.}
The analysis of pathways to non-sequential double ionization of atoms
in strong fields \cite{eckhardt01pra1,eckhardt01,eckhardt01epl} 
allows us to propose a 1+1 dimensional model of 
the process. In the present publication we analyze the model in the case 
of a static electric field. 
The Wannier threshold law for simultaneous electron escape in the presence 
of the static field is derived and tested numerically. We also identify 
separatrices that divide the phase space of the system into areas of
simultaneous electron escapes and double escapes with a multiple 
revisit of the nucleus by one of the electrons. In classical mechanics
these different trajectories contribute independently to the double ionization 
process. In quantum mechanics one may expect interesting coherence effects
resulting from interference of the different paths.

The model is considered classically and for a static external field only. 
However, quantum calculations of double ionization
in strong laser field \cite{corkum93,kulander93,silap00} 
including: electron tunneling, rescattering and 
subsequent double escape can be performed numerically very efficiently 
within the model. As we have
mentioned in the introduction the crucial advantage of the model over
the well known aligned electron model 
\cite{align1,align2,align3,align4} is that it does not forbid 
symmetric simultaneous escapes of electrons what is observed experimentally 
and what can not be described by the aligned electron model due to 
an overestimation of the Coulomb repulsion.

This work was partially supported by the Alexander von Humboldt 
Foundation, the Deutsche Forschungsgemeinschaft and the KBN through grant 
(KS) PBZ-MIN-008/P03/2030.

\end{document}